# B.A.T.Mobile: Leveraging Mobility Control Knowledge for Efficient Routing in Mobile Robotic Networks

**Benjamin Sliwa, Daniel Behnke, Christoph Ide and Christian Wietfeld**
Communication Networks Institute
TU Dortmund University
44227 Dortmund, Germany
e-mail: {Benjamin.Sliwa, Daniel.Behnke, Christoph.Ide, Christian.Wietfeld}@tu-dortmund.de

***Abstract*—Efficient routing is one of the key challenges of wireless networking for unmanned aerial vehicles (UAVs) due to dynamically changing channel and network topology characteristics. Various well known mobile-ad-hoc routing protocols, such as AODV, OLSR and B.A.T.M.A.N. have been proposed to allow for proactive and reactive routing decisions. In this paper, we present a novel approach which leverages application layer knowledge derived from mobility control algorithms guiding the behavior of UAVs to fulfill a dedicated task. Thereby a prediction of future trajectories of the UAVs can be integrated with the routing protocol to avoid unexpected route breaks and packet loss. The proposed extension of the B.A.T.M.A.N. routing protocol by a mobility prediction component – called B.A.T.Mobile – has shown to be very effective to realize this concept. The results of in-depth simulation studies show that the proposed protocol reaches a distinct higher availability compared to the established approaches and shows robust behavior even in challenging channel conditions.**

## I. Introduction

Mobile robotic networks are an important subset of mobile ad-hoc networks (MANETs) and form a class for a wide range of different network types. Applications range from dynamic traffic management in the field of vehicular ad-hoc networks (VANETs) to maintaining robust swarm communication for Unmanned Aerial Vehicles (UAVs) exploring hazardous areas. The provision of reliable end-to-end communication in this kind of networks is a challenging topic due to the high relative mobility. Established routing protocols can barely cope with the frequently changing network and fail to adopt to the new channel and topology conditions. This issue is widely known and has been onesidedly addressed from two different perspectives: a mobility-centric view and a routing-centric view. In this paper, we combine these approaches to enhance the overall routing performance and the stability of communication paths in low altitude UAV networks. Application layer mobility control data is used to predict future node positions. This information is then used to enable a forward-looking routing approach to optimize the packet forwarding process. We analyze the system behaviour with multiple mobility algorithms (see Fig. 1), which are described in Section IV-A. Our simulation setup has been published as an Open Source framework and is described in [1]. The remainder of this paper is structured as follows: after discussing the related work, we present the system model of our solution approch, which contains the novel prediction method and our proposed routing protocol as subchapters. In the next section we describe the used mobility algorithms, the traffic model and our simulation environment. Finally, detailed results of multiple simulation evaluations are presented, which compare our proposal to existing approaches. The results show the high efficiency of our proposed methods and prove their suitability for highly dynamic mobile robotic networks.

## II. Related Work

Challenges of wireless communications with UAVs are discussed in [2]. The high mobility is one important challenge for these networks. Recent approaches try to solve the issue by optimizing the mobile behavior but neglect the influence on the routing. A classification about different methods of using mobility information to enhance the routing process is given in [3]. A common approach is the estimation of link expiration times in order to forward packets via the neighbor with the highest link-availability [4]. Most approaches assume constant movement vectors, which cannot be fulfilled for real-world UAV exploration tasks. The usage of movement traces in order to predict future trajectories is evaluated in [5] in the field of VANETs. The authors of [6] use mobility prediction with a geographic routing protocol to avoid routing voids. Knowledge about planned trajectories is used by the Trajectory Aware Geographical (TAG) routing protocol in [7] in order to avoid link-breaks in Cognitive Radio Ad Hoc Networks (CRAHNs). The authors are able to decrease the end-to-end delay significantly. The authors of [8] present Mobility and Load aware OLSR (ML-OLSR) as an extension to the well-known Optimized Link State Routing (OLSR) protocol and uses information about the nodes movement direction to optimize the multipoint relay selection. The results show an improved packet delivery ratio (PDR). An overview about bio-inspired routing protocols is given in [9] and the 'Mobility aware-Termite' algorithm is presented, which uses the node's position history to improve the path availability in combination with pheromone-based routing. It is able to outperform established reactive protocols in terms of throughput and delay. However the node-distances are estimated from the received signal strength, which will cause frequent positioning errors in real-world channel conditions. The discussed approaches indicate the scientific interest and the relevance of the topic. Existing methods are often based on position histories and movement directions but do not interact with the mobility control layer in order to optimze the prediction accuracy.

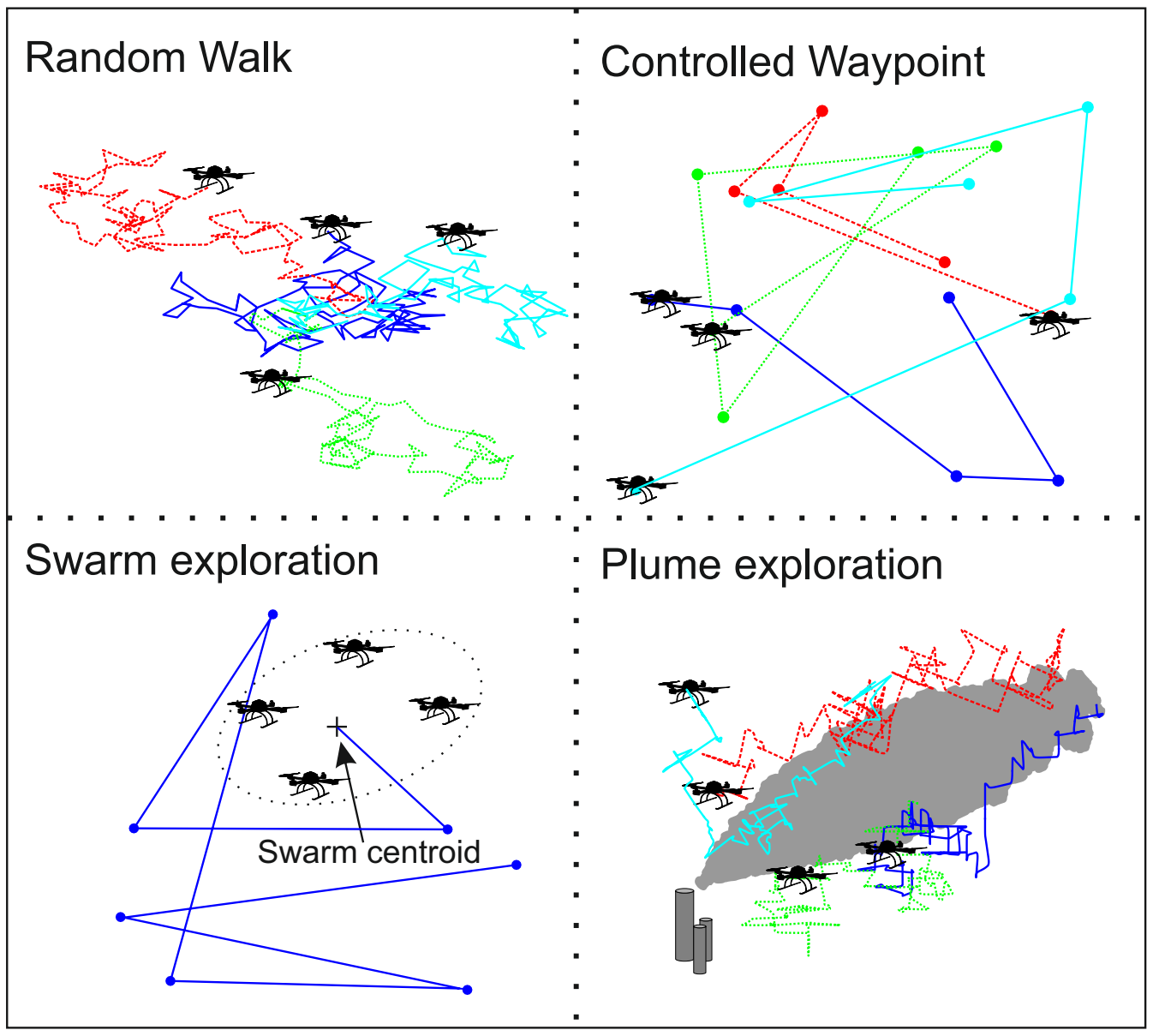

Fig. 1. Typical trajectories for different mobility algorithms

## III. Cross-layer solution approach

Our proposed system model is illustrated in Fig. 2. The UAV control software implements the mobility algorithms and acts as a database for the agent's current mobility information, which is used by our proposed multifactoral prediction method to determine the future agent trajectory. For the routing process the basic Better Approach To Mobile Adhoc Networking (B.A.T.M.A.N.) [10] routing protocol is used to handle the exchange of routing messages between the agents because of its stigmergic approach, which provides a sophisticated way of distributing path quality information through the network. The forwarding decision is performed by our novel forward-looking metric and utilizes the results of the mobility prediction process in order to optimize the neighbor selection, leading to an improved packet delivery ratio. In the following subchapters we will give a detailled description of the novel mobility prediction algorithm and present B.A.T.Mobile as an extension to the B.A.T.M.A.N. protocol.

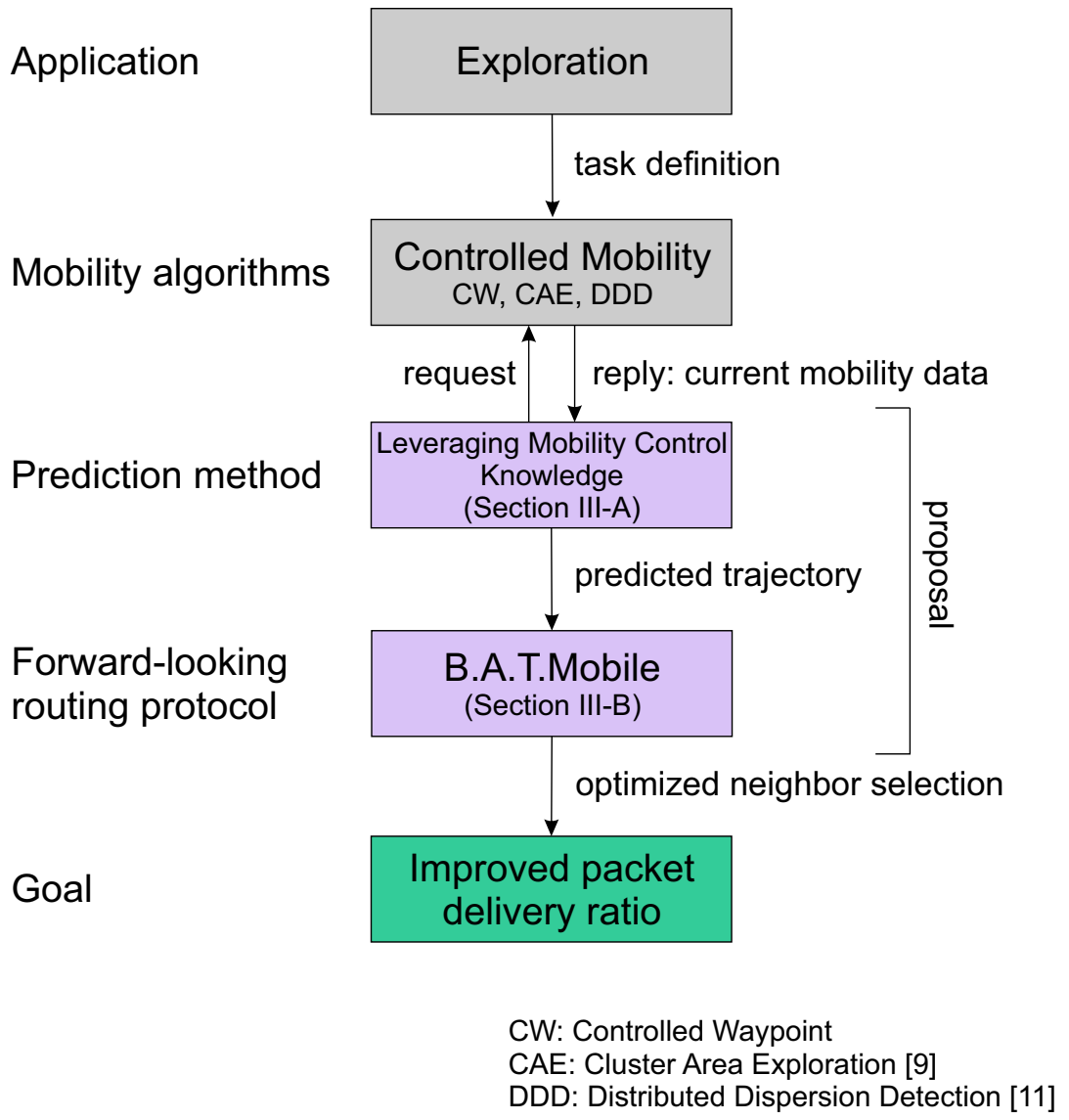

Fig. 2. Cross-layer solution approach: utilizing predicted mobility information for the routing decisions

### *A. Leveraging mobility control knowledge for node trajectory prediction*

The stepwise prediction process uses different kinds of mobility information and is illustrated for an example agent in Fig. 3. A new position estimate is calculated for each next step until the final iteration $N_p$ is reached. Within each iteration $i$, the prediction method with the assumed highest precision is selected from all available methods to perform the calculation for the next position estimation step. Since the prediction process is iterative, previously predicted position values can also be used by the extrapolation process. Fig. 4 summarizes the prediction algorithm as a flow chart. The *steering vector* $\vec{\sigma}_i$ represents the vector to the desired vehicle position $\vec{P}_{i+1}$ in the next update step $i+1$ and is calculated by the vehicle control software. Since it is a weighted superposition of individual steerings taking into account exploration, collision avoidance and swarm coherence, it cannot be predicted itself and is only available in the first step of the prediction process. For the prediction of the next step, $\vec{\sigma}_i$ needs to be scaled from the update interval $\Delta t_u$ of the mobility algorithm to the actual time difference $t_{i+1} - t_i$ (see Eq. 1) .

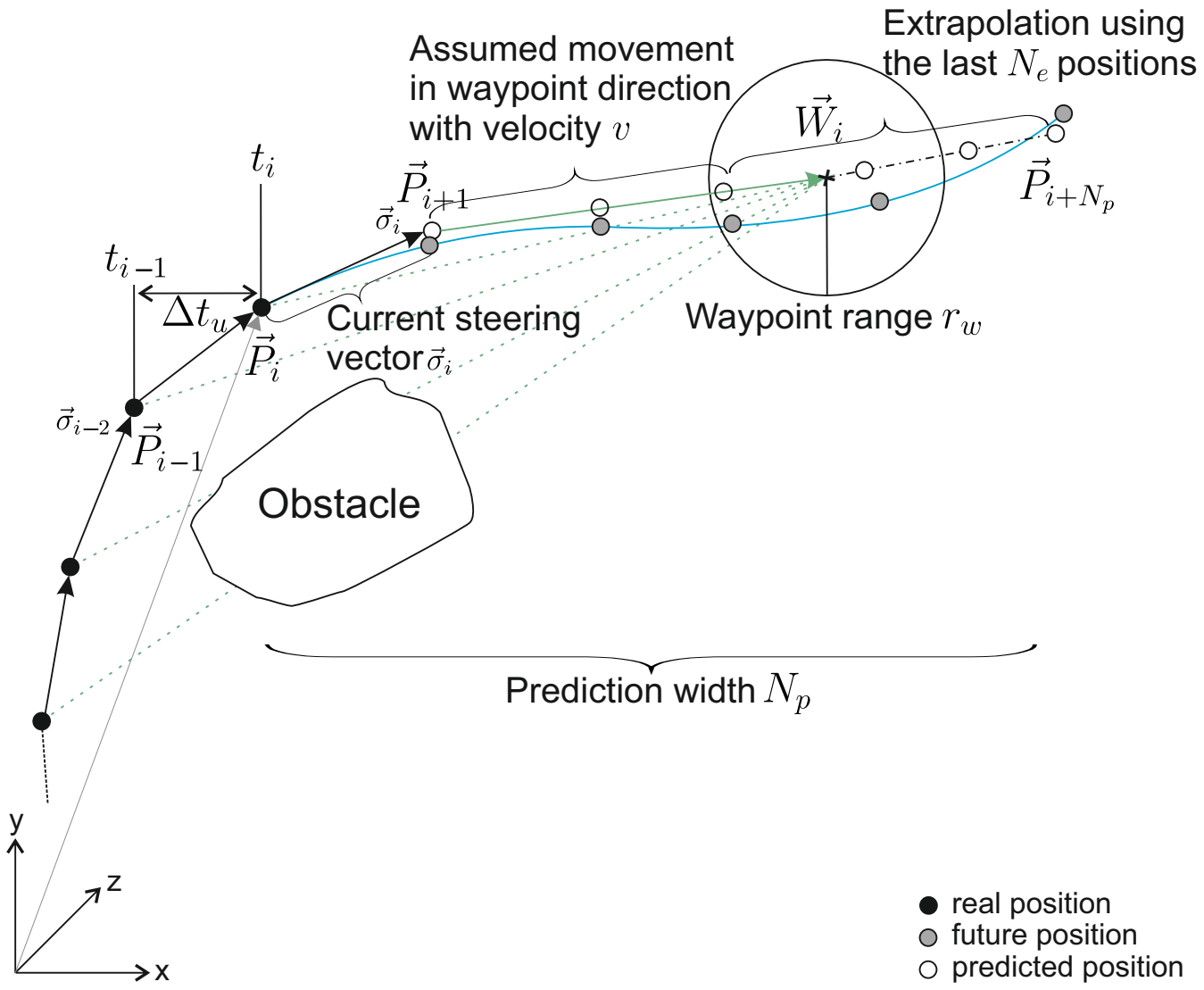

Fig. 3. Application of the predicion method for an example agent

$$\vec{P}'_{i+1} = \vec{P}_i + (t_{i+1} - t_i)\,\frac{\vec{\sigma}_i}{\Delta t_u} \tag{1}$$

Individual vehicle waypoints act as orientation points and indicate the current movement direction of the agent. Depending on the mobility algorithm the frequency of waypoint changes is variable. Eq. 2 is used to determine the next position on the line-of-sight to the current waypoint $\vec{W}$ with a defined velocity $v$.

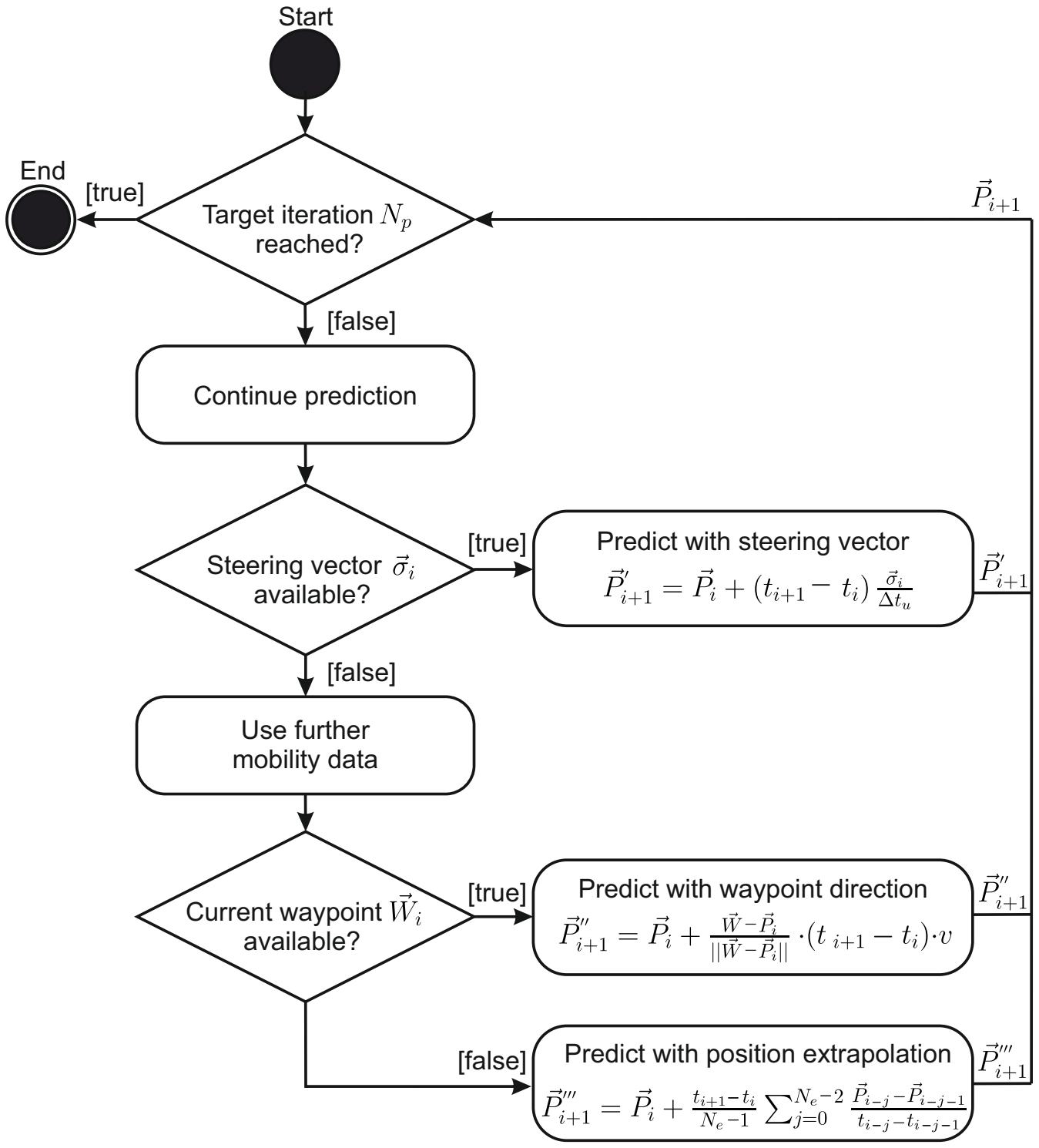

Fig. 4. Flow Chart for the iterative prediction process for B.A.T.Mobile: in each iteration the estimation of the next position is determined by the prediction method with the highest available precision

$$\vec{P}''_{i+1} = \vec{P}_i + \frac{\vec{W} - \vec{P}_i}{||\vec{W} - \vec{P}_i||} \cdot (t_{i+1} - t_i) \cdot v \tag{2}$$

The distance of the predicted position to the current waypoint is contineously determined to check if the agent is inside the waypoint range $r_w$ and can be considered "reached". A change of the current waypoint is performed after a waypoint has been reached and further waypoints are available. If no waypoints are remaining, this prediction method can no longer be used. As a fallback solution, the extrapolation prediction method is used, if no other information is available. It calculates the average movement vector from the last $N_e$ positions as shown in Eq. 3.

$$\vec{P}'''_{i+1} = \vec{P}_i + \frac{t_{i+1} - t_i}{N_e - 1} \sum_{j=0}^{N_e-2} \frac{\vec{P}_{i-j} - \vec{P}_{i-j-1}}{t_{i-j} - t_{i-j-1}} \tag{3}$$

### *B. Predictive routing with B.A.T.Mobile*

The main task of the routing process is the choice of a forwarder node from the available neighbors for a given destination. Most established protocols only maintain a single path for each destination in their routing tables, limiting the capabilities for proactive avoidance of path losses in highly dynamic networks. Moreover the usage of simple decision metrics (e.g. hop count) leads to a high frequency of route switches. To overcome these issues, every node $N$ uses a *Neighbor Ranking* (Fig. 5) for each destination $D$, which contains all neighbor nodes of $N$ and a score as an indicator about their suitability to be the next hop on the path $D$. The routing decision is simplified to selecting the neighbor with highest score from the ranking. We use B.A.T.M.A.N.

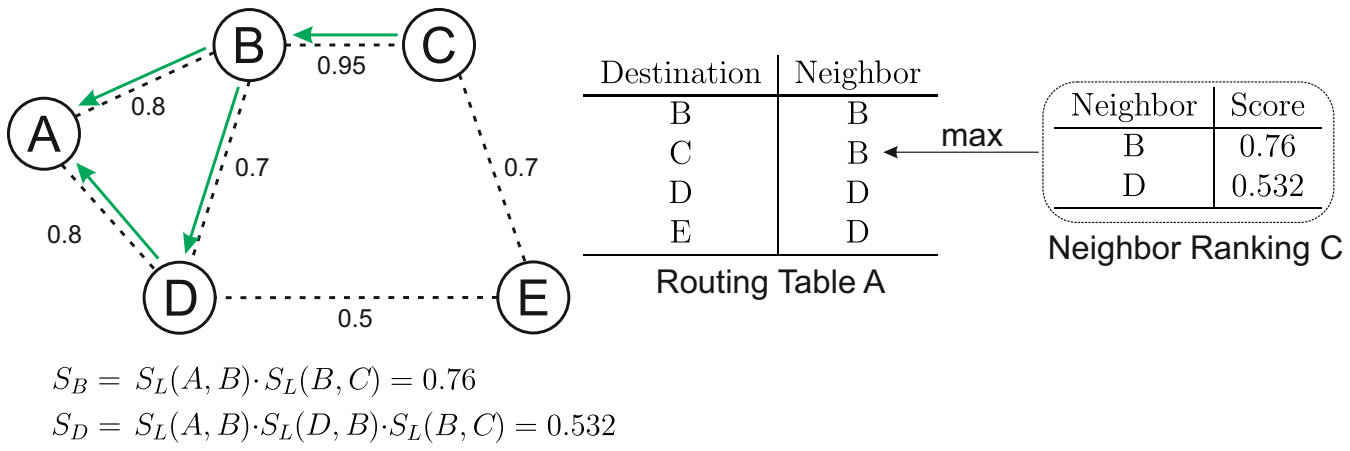

Fig. 5. Score-based maintenance of multiple routing paths using neighbor rankings in an example network. Node $A$ calculates the neighbor scores for the destination $C$ from the received messages.

as the basic routing protocol to forward the required mobility information with its Originator Messages (OGMs). All messages are extended with the current forwarder position $\vec{P}$, the predicted forwarder position $\vec{P'}$ and the current path score $S$. If a node generates a broadcast message, it initializes $S$ with the maximum value 1 and sets both position values to its own positions. On reception of a routing packet, the new path score $S'$ is calculated by multiplicating the received path score $S$ with the link score $S_L$ to the forwarder node as shown in Eq. 4. This method implicitely punishes paths with high hopcounts. $S'$ and the position values of the receiver node are then used to update the routing message and forward it according to the rules of the basic protocol.

$$S' = S \cdot S_L \tag{4}$$

For the link score $S_L$ the calculation is based on the distance $d$ to the forwarder node, which is set in relation to a maximum distance $d_{max}$. The latter is obtained with a defined channel model for a desired minimal received signal strength $P_{e,min}$. Depending on the potential dynamics of the network topology in the considered scenario, the parameter $\alpha$ is used to control the tradeoff between the influence of absolute distances and relative mobility on the total score.

$$S_L = \min\left[1 - \left(\frac{d}{d_{max}}\right)^{\alpha}, 1 - \left(\frac{d'}{d_{max}}\right)^{\alpha}\right] + p_{trend} \tag{5}$$

The relative agent mobility is taken into account by Eq. 6, which correlates the predicted development of the distance to the forwarder with the maximal possible distance using the vehicle's step width $d_{step}$. The resulting value range is defined by the parameter $p_{trend,max}$.

$$p_{trend} = \begin{cases} 0 & : N_P = 0 \\ \frac{d'-d}{2 \cdot d_{step} \cdot N_P} \cdot p_{trend,max} & : \text{else} \end{cases} \tag{6}$$

Since routing packets are usually sent with best-effort delivery, packet losses occur and may cause false routing decisions.

B.A.T.Mobile addresses this issue with the introduction of the *Neighbor Score Buffer* B (see Fig. 6). New path scores are only used to update the current score candidate $S_C$. The timer $t_u$ controls the update phase, which sets the value of $S_C$ to the score of the best received path to the destination. After the timeout occurence $S_C$ is shifted to the neighbor score buffer and is then reset to zero. The resulting neighbor score is calculated as the mean of the buffer and assigned to the forwarder node in the neighbor ranking.

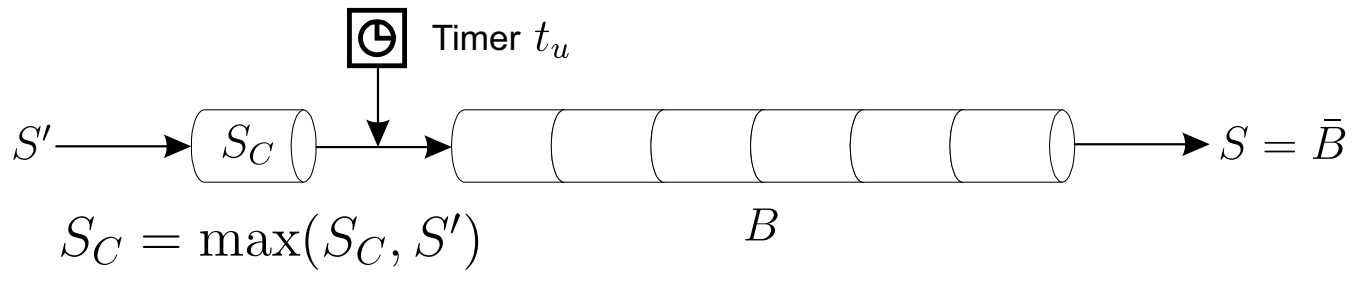

Fig. 6. Compensation of lost routing packets through score buffering

## IV. Simulation-based System Model

In this section, the simulation-based system model that is used for the performance evaluation is presented. It consists of the description of the mobility algorithms, the actual routing simulation with scenario definition and the data traffic model.

### A. Mobility algorithms

We use the *Random walk* mobility model in order to determine a lower bound for the benefit through using the prediction method. In each iteration the movement vector is determined randomly. For the general performance evaluation of our proposed methods, we use a *Controlled Waypoint* algorithm, which uses a trajectory of multiple random positions. The random positioning causes frequent changes of the network topology, leading to challenging situations for the routing protocols. In contrast to the well-known *Random Waypoint* algorithm, future waypoints are known from the beginning and can therefore be utilized by the prediction method. Implications for real-life UAV applications are derived from analyzing the routing behaviour using two different exploration algorithms. *Cluster Area Exploration (CAE)* [11] performs a swarm-based exploration by selecting random waypoints for the swarm centroid. The algorithm is used in combination with *Communication Aware Potential Fields (CAPF)* [12], which maintains the swarm coherence and builds up a chain structure to the base station at runtime. The *Distributed Dispersion Detection (DDD)* [13] algorithm is used for plume detection and is able to maintain the swarm coherence on its own. For the exploration task, the swarm uses a mesh structure with high relative agent mobility.

### B. Traffic model

Our reference scenario is defined by a swarm of autonomous agents exploring a mission area. One of the agents is randomly selected to continously stream User Datagram Protocol (UDP) video data to the base station, which is centered inside the territory. Telemetry information is periodically broadcasted by all agents in order to keep the swarm coherence and collision avoidance steerings updated with recent data.

### C. Routing simulation with OMNeT++/INETMANET

We use the discrete event-based simulation environment OMNeT++ [14] and its INETMANET framework for the evaluation of the routing protocols. For the integration of application layer mobility data, INETMANET has been enhanced with a dedicated location service, which makes those information available for network layer routing protocols. Additionally a new base module *"GeoAssistedRoutingBase"* has been added in order to provide score-based routing tables and our novel path score metric to further protocols. The simulation parameters for the reference scenario are defined in Tab. I. Deviations from the default assignment are explicitely marked where they are required.

TABLE I
Simulation Parameters for the reference scenario in OMNeT++/INETMANET

| **Simulation parameter** | **Value** |
|---|---|
| Mission area | 500m x 500m x 250m |
| Number of agents | 10 |
| Mobility model | Controlled Waypoint |
| Velocity $v$ | 50 km/h |
| Channel model | Friis, Nakagami ($\gamma = 2.75$) |
| Videostream bitrate | 2 Mbit/s |
| Telemetry broadcast interval | 250 ms |
| Telemetry packet size | 1000 Byte |
| OGM broadcast interval (B.A.T.M.A.N.) | 0.5 s |
| HELLO interval (OLSR) | 0.5 s |
| Topology Control (TC) interval (OLSR) | 1 s |
| Medium Access Control (MAC) layer | IEEE802.11g |
| Transport layer protocol | UDP |
| UDP Maximum Transmission Unit (MTU) | 1460 Byte |
| Transmission power | 100 mW |
| Carrier frequency | 2.4 GHz |
| Receiver sensitivity | -83 dBm |
| Simulation time per run | 300 s |
| Number of simulation runs | 50 |
| GNSS positioning error | 0, [0,120] m |
| Neighbor Score Buffer size | 8 |
| Mobility update interval $\Delta t_u$ | 250 ms |
| Extrapolation data size $N_e$ | 5 |
| Prediction width $N_p$ | 15, [0, 30] |
| Grade of relative mobility $\alpha$ | 7 |
| Maximum path trend $p_{trend,max}$ | 0.1 |

## V. Results

In this section we present the results achieved with our proposed protocol extension. We consider the PDR of the video stream to the base station as our main key performance indicator.

### A. Comparison with established routing protocols

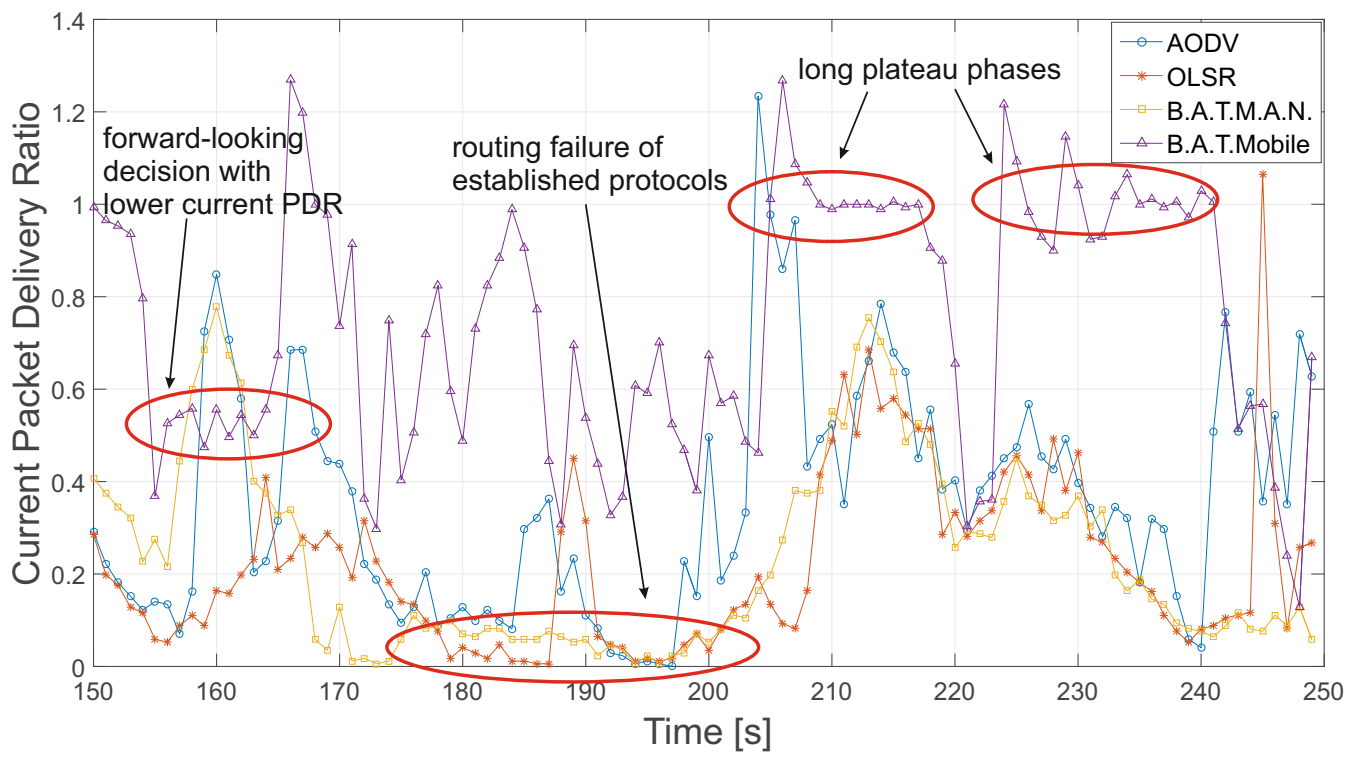

Fig. 7. Example temporal behaviour comparison of B.A.T.Mobile with established routing protocols in urban environments

We compare the results of B.A.T.Mobile with B.A.T.M.A.N. and the established routing protocols OLSR [15] and Ad-hoc On-demand Distance Vector (AODV) [16] for different mobility algorithms. General characteristics of the routing behaviour can be identified analyzing the time behaviour, which is exemplarily shown in Fig. 7. Using the proposed routing metric, B.A.T.Mobile avoids packet losses and and is able to maintain high PDR-values for longer time intervals compared to its competitors. On rare occasions the forward-looking approach leads to situations, where the current PDR is temporarily below the value of the established protocols. Each routing decision is a trade-off between the current and the predicted network state, therefore it is sometimes required to spare current PDR peaks in order to avoid drops in the near future. Fig. 8 shows the statistical results for a scenario with *Controlled Waypoint* mobility in two application scenarios. We use a Friis channel model for rural scenarios and a Nakagami channel model $(m = 2)$ for urban applications. B.A.T.Mobile outperforms the established protocols significantly, while B.A.T.M.A.N. shows the lowest PDR in both scenarios. AODV achieves the highest PDR value for the established protocols and is fitting for mobile applications with low packet loss probabilities due to its reactive nature. Packet losses are more probable in urban scenarios, causing lower PDR values for all protocols. AODV suffers heavily from the loss of route request (RREQ) and route reply (RREP) packets, which results in a high PDR decrease. The impact of the channel conditions on the overall routing performance is much lower for B.A.T.Mobile than for its competitors. The path score is contained in all routing messages and acts an indicator for the quality of the whole path to the message originator. The effect of packet losses is furthermore reduced by the neighbor score buffer. Fig. 9 shows the resulting PDR values of B.A.T.Mobile and B.A.T.M.A.N. for different mobility algorithms. As expected, the *Random Walk* routing performance cannot be significantly improved using predictive methods, although the score buffering slightly increases the PDR. For the Swarm exploration algorithm the general PDR is relatively low due to the chain structure, which contains many single point of failure links. The best routing performance is achieved with the plume exploration algorithm. B.A.T.Mobile highly benefits from the inherent path choice possibilities of the dynamical mesh structure and is able to enhance the mean PDR value above 90%.

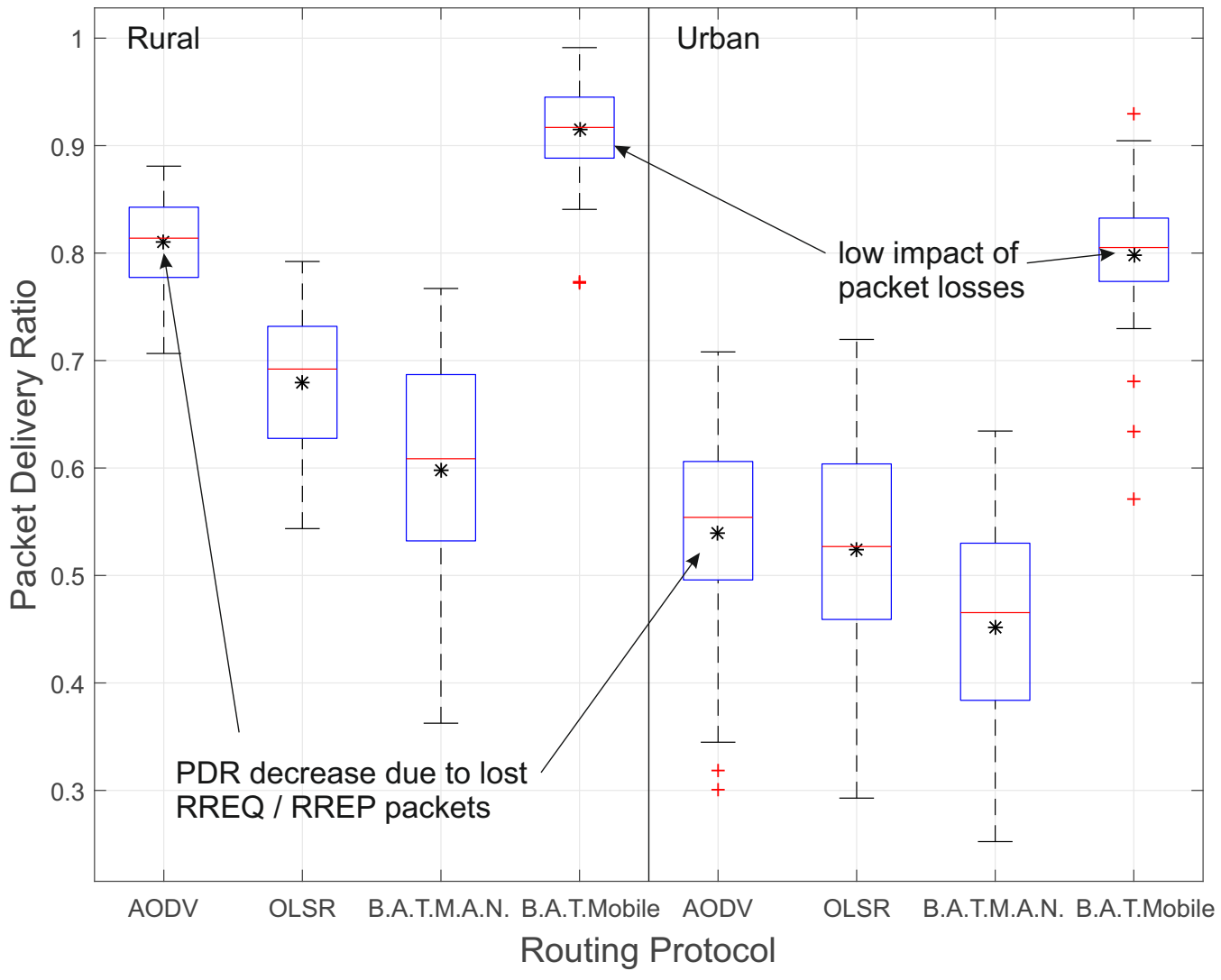

Fig. 8. Comparison of the routing protocols for *Controlled Waypoint* mobility in two channel models

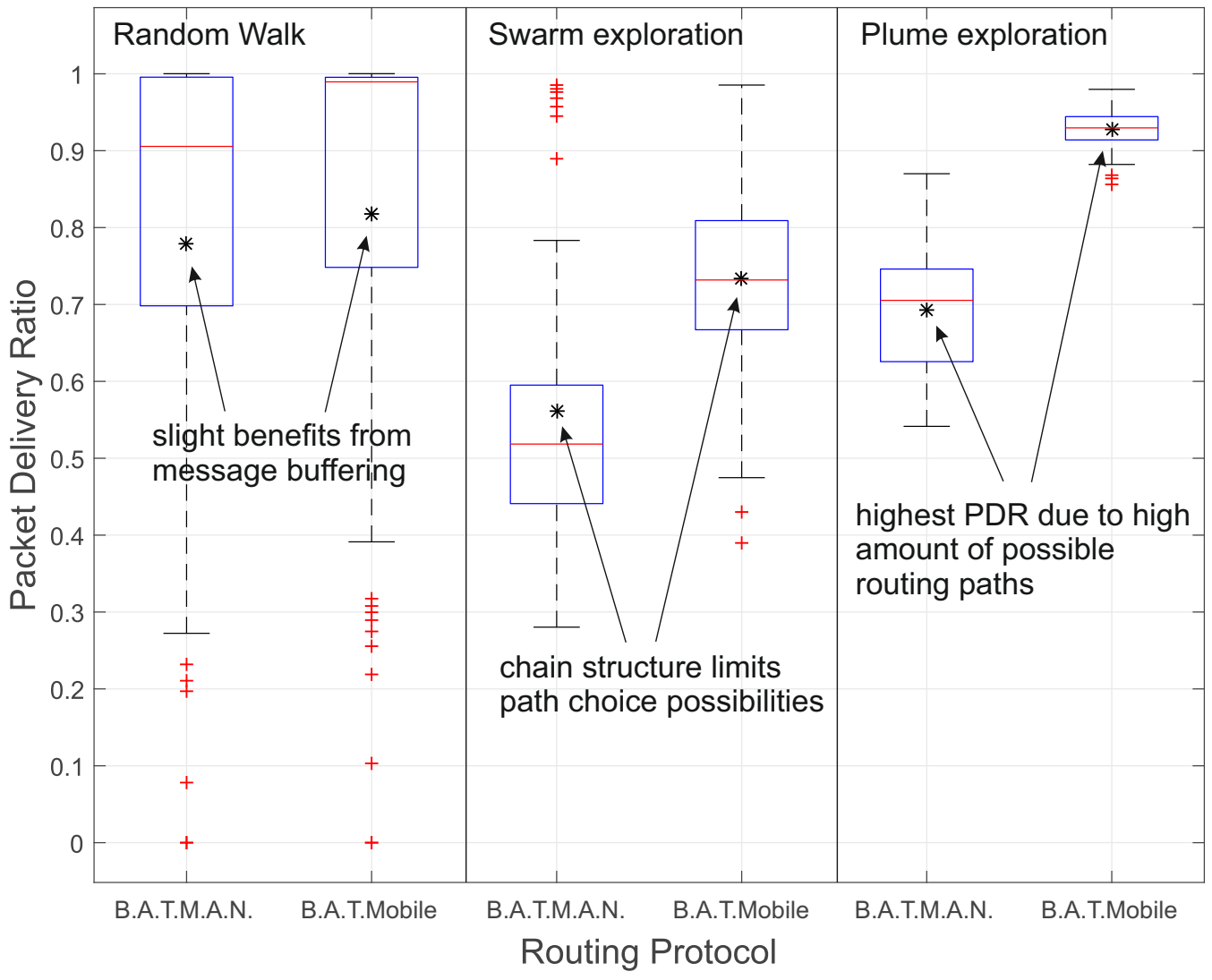

Fig. 9. Performance of B.A.T.Mobile in relation to B.A.T.M.A.N. in different controlled mobility scenarios (Rural)

## B. Parameterization

In order to evaluate the influence of individual mobility information on the total routing performance, we compare multiple parameter configurations. Fig. 10 shows the resulting

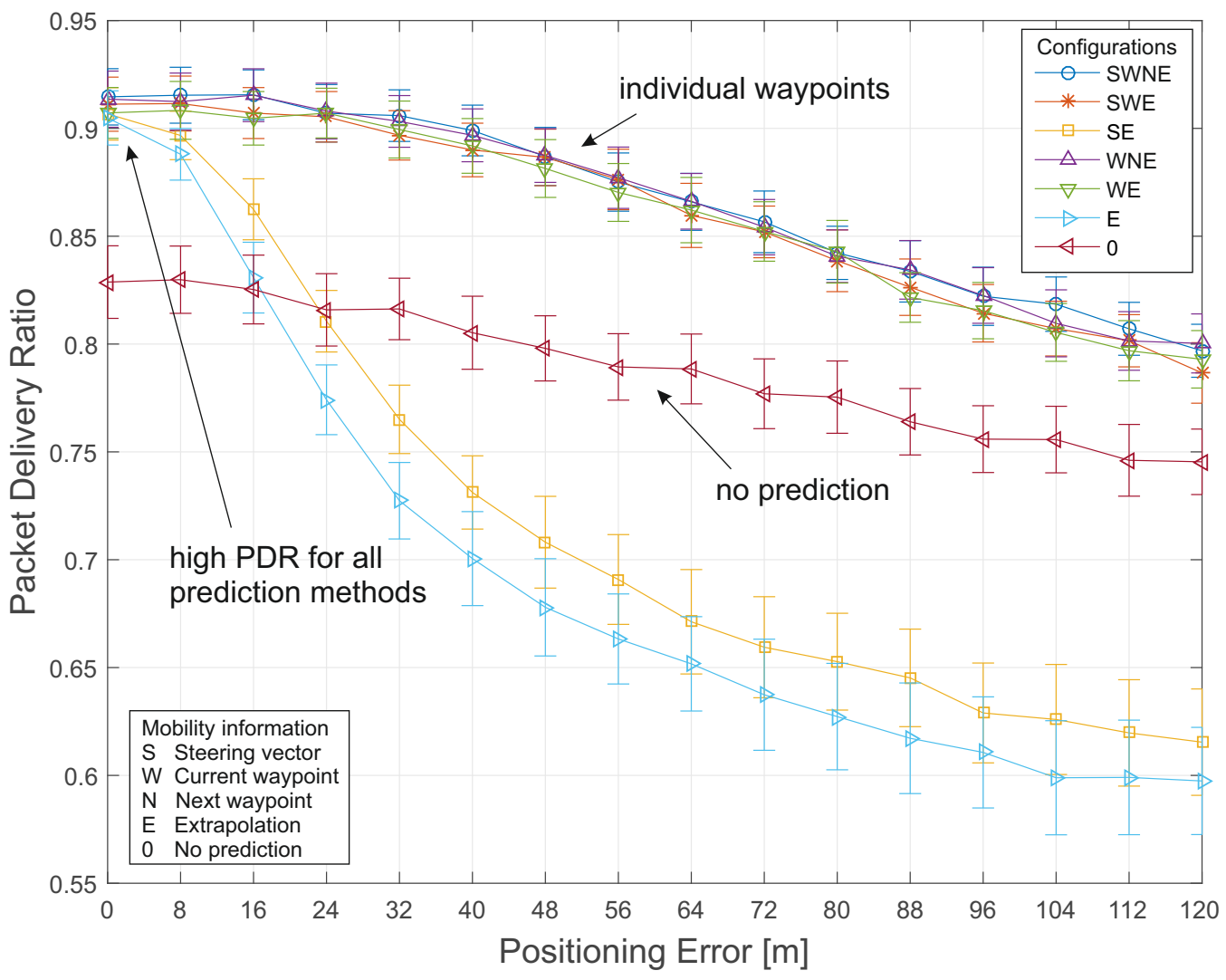

Fig. 10. Sensitivity analysis: Impact of positioning errors on the routing performance for different mobility control information configurations

PDR values for the different configurations depending on the GNSS positioning accuracy. Generally we can identify three effects:

1) If the positioning error range is very low, all configurations achieve nearly equally high PDR values.

2) Individual waypoints act as static orientation points and reduce the influence of positioning errors on the overall routing performance. Further mobility information only bring minor benefits.
3) The extrapolation prediction method fails, if the positioning error is higher than a threshold. Enhancements can be achieved by integrating the *steering vector*.

The prediction width is one of the key parameters for the performance of the routing decisions. It is depending on the intensity of the dynamic of the network topology and influenced by the mobility pattern of the agents. Fig. 11 shows the selection of the optimal prediction width for multiple movement speed parameterizations. The possible benefit of using predictive methods is proportional to the velocity. For lower speeds the general PDR is high, thus limiting the possible space for further improvements. For higher speed values the PDR can be highly increased by the prediction but the dependency to an optimal parameter choice is also intensified.

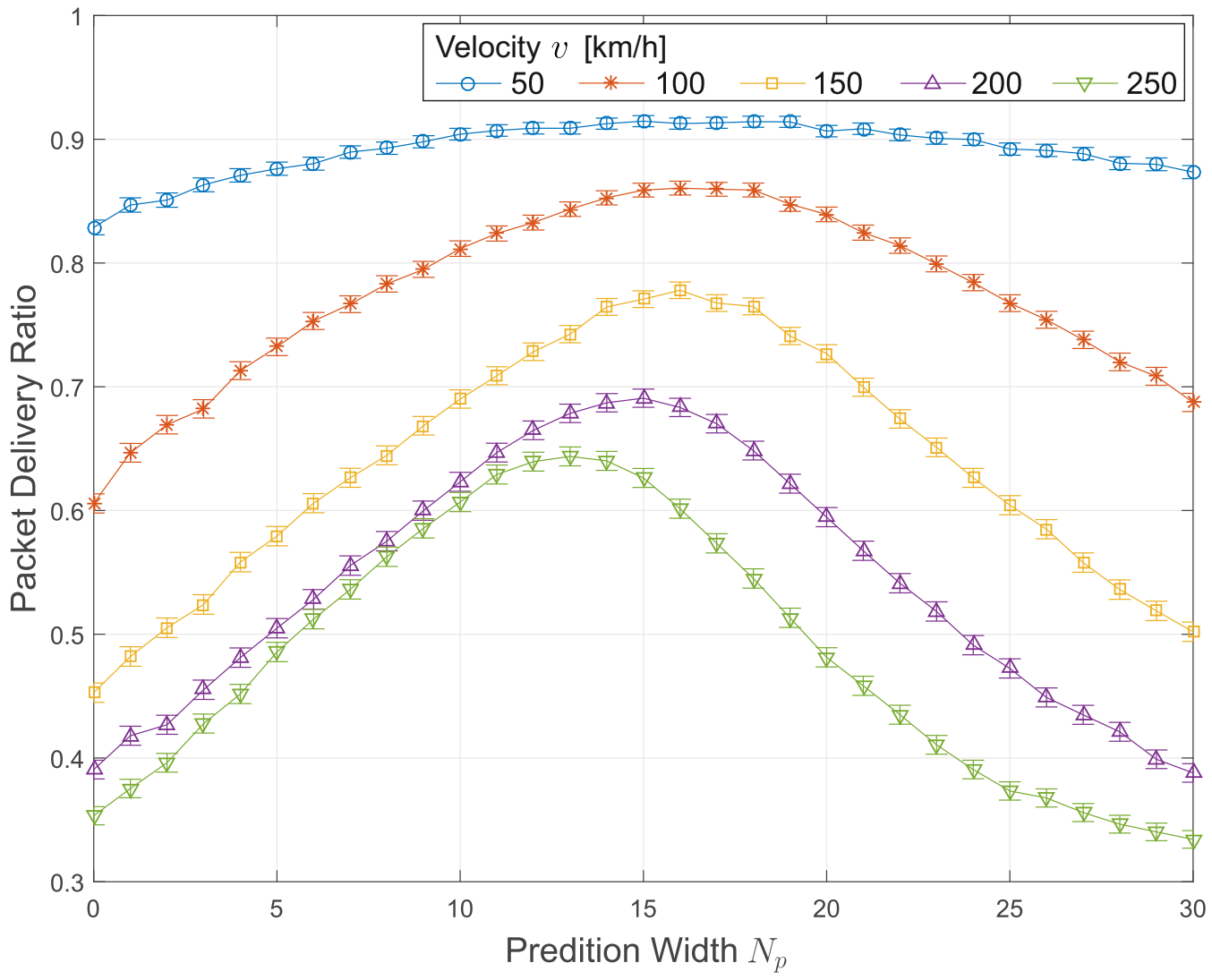

Fig. 11. Choice of the prediction width depending on the vehicle's velocity

## VI. Conclusion

In this paper, we presented B.A.T.Mobile as a novel cross-layer approach to avoid losses of communication paths in highly dynamic robotic networks. The proposed protocol uses a novel prediction algorithm in order to optimize the selection of forwarder nodes by taking the relative agent mobility into account. We compared our proposal with multiple established routing protocols and demonstrated its superiority in different mobility scenarios. The results of the simulation proved the ability of B.A.T.Mobile to provide reliable communication even under challenging environmental conditions. Furthermore we showed that the utilization of application layer mobility information can significantly enhance the overall routing performance. In the future, we will investigate the suitability of B.A.T.Mobile to be used in the field of VANETs applications. Additionally, we want to evaluate the usage of our prediction method and the path score routing metric with more routing protocols.

## Acknowledgment

Part of the work on this paper has been supported by Deutsche Forschungsgemeinschaft (DFG) within the Collaborative Research Center SFB 876 "Providing Information by Resource-Constrained Analysis", project B4 and has received funding from the European Union Seventh Framework Programme (FP7/2007-2013) under grant agreement n°607832 (project SecInCoRe). The text reflects the authors' views. The European Commission is not liable for any use that may be made of the information contained therein. For further information see http://www.secincore.eu/.